\documentclass[aps,pre,reprint,superscriptaddress]{revtex4-1}

\usepackage{graphicx}
\usepackage{float}
\usepackage{times}

\newcommand{\Valles}{Toni Vall\`es-Catal\`a}
\newcommand{\Massucci}{Francesco A. Massucci}
\newcommand{\Guimera}{Roger Guimer\`a}
\newcommand{\Sales}{Marta Sales-Pardo}
\newcommand{\URV}{Departament d'Enginyeria Qu\'{\i}mica, Universitat Rovira i Virgili, 43007 Tarragona, Catalonia}
\newcommand{\ICREA}{Instituci\'o Catalana de Recerca i Estudis Avan\c{c}ats (ICREA), Barcelona 08010, Catalonia}

\newcommand{\M}{\cal M}
\renewcommand{\P}{{\cal P}}
\renewcommand{\O}{\cal O}

\begin{document}

\title{Multilayer stochastic block models reveal the multilayer structure of complex networks}

\author{\Valles}
\affiliation{\URV}
\author{\Massucci}
\affiliation{\URV}
\author{\Guimera}
\email[Corresponding author: ]{roger.guimera@urv.cat}
\affiliation{\ICREA}
\affiliation{\URV}
\author{\Sales}
\email[Corresponding author: ]{marta.sales@urv.cat}
\affiliation{\URV}

\begin{abstract}
In complex systems, the network of interactions we observe between system's components is the aggregate of the interactions that occur through different mechanisms or layers. Recent studies reveal that the existence of multiple interaction layers can have a dramatic impact in the dynamical processes occurring on these systems. However, these studies assume that the interactions between systems components in each one of the layers are known, while typically for real-world systems we do not have that information. 
Here, we address the issue of uncovering the different interaction layers from aggregate data by introducing multilayer stochastic block models (SBMs), a generalization of single-layer SBMs that considers different mechanisms of layer aggregation. First, we find the complete probabilistic solution to the problem of finding the optimal multilayer SBM for a given aggregate observed network. Because this solution is computationally intractable, we propose an approximation that enables us to verify that multilayer SBMs are more predictive of network structure in real-world complex systems.  
\end{abstract}


\maketitle


The development of tools for the analysis of real-world complex networks
%
%
has significantly advanced our understanding of complex systems in fields as diverse as molecular and cell biology~\cite{barabasi04}, neuroscience~\cite{bullmore09}, biomedicine~\cite{barabasi11,csermely13}, ecology~\cite{thompson12,rohr14}, economics~\cite{schweitzer09}, and sociology~\cite{borgatti09}. One of the main successes of the network approach has been to unravel the relationship between the modular organization of interactions within a complex system \cite{newman11}, and the function and temporal evolution of the system~\cite{guimera05a,arenas06,guimera07,ahn10}. As a result, a large body of research has been devoted to the detection of the modular structure (or community structure) of complex networks, that is, to the division of the nodes of the network into densely connected subgroups~\cite{fortunato10}.

Stochastic block models (SBMs) \cite{white76,holland83,nowicki01} are a class of probabilistic generative network models that provide a more general description of the (mesoscopic) group structure of real-world networks than modular models. In SBMs, nodes are assumed to belong to groups and connect to each other with probabilities that depend only on their group memberships. The simple mathematical form of SBMs has enabled not only the identification of generalized community structures in networks \cite{nowicki01,karrer11,decelle11,schmidt13,peixoto13,peixoto14,peixoto14b,larremore14,aicher14,yan14}, but also to make network inference a predictive tool to detect missing and spurious links in empirical network data~\cite{guimera09}, to predict human decisions \cite{guimera11,guimera12} and the appearance of conflict in work teams~\cite{roviraasenjo13}, and for the identification of unknown interactions between drugs~\cite{guimera13}.
 
While these approaches have pushed forward our understanding of complex network structure, a limitation is that they rely on the premise that there is a single mechanism that describes the connectivity of the network, even though we know that real-world networks are often the result of processes occurring on different ``layers'' (for example, social networks comprise relationships that arise in the familiar layer, and others that arise in the professional layer)~\cite{kivela14}. Moreover, it is increasingly clear that the multilayer structure of complex networks can have a dramatic impact on the dynamical processes that take place on them~\cite{morris12,radicchi13,gomez13,dedomenico13,dedomenico14}. Unfortunately, we often lack information about the different layers of interaction and can only observe projections of these multilayer interactions into an aggregate network in which all links are equivalent. 


Here, we precisely address the problem of unraveling the underlying multilayer structure in real-world networks. To do so, we first introduce the family of multilayer SBMs that generalizes single-layer SBMs to situations where links arise in different layers and are aggregated through different mechanisms. Although there have been proposals to extend the concept of modularity to multilayer networks~\cite{mucha10}, ours represents a pioneering attempt to extend generative group-based models to multilayer systems, and to study those models rigorously using tools from statistical physics.

 Second, we give the probabilistically complete solution to the problem of inferring the optimal multilayer SBM for a given aggregate network. Because this solution is computationally intractable, we propose an approximation which enables us to objectively address the question of whether an observed network is likely to be the projection of multiple layers. Our results suggest that many real-world networks are indeed projections.




\section{Multilayer stochastic block models}

\begin{figure}
  \includegraphics*[width=.6\columnwidth]{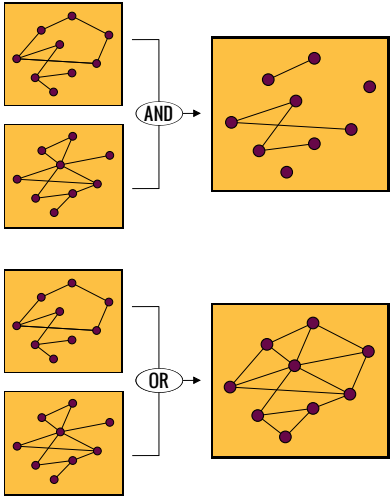}
  \caption{In aggregated multilayer networks, several networks with the same nodes but different links are combined into an observed network $A^\mathcal{O}$ that has no information about the original layers. Two-layer networks can be aggregated using an AND combination of the layers or an OR combination of the layers. {\bf (a)} The AND aggregation has link $A^\mathcal{O}_{ij}=1$ if, and only if, $i$ and $j$ are connected in both layers.
  {\bf (b)} The OR aggregation has link $A^\mathcal{O}_{ij}=1$ if, and only if, $i$ and $j$ are connected in at least one layer.
  \label{f.schema}}
\end{figure}

In our approach, nodes interact in different layers. In each one of these layers $\ell=1,\dots,L$ we define a SBM as follows: each node $i$ belongs to a specific group $\sigma_i^{\ell}$, and links between pairs of nodes belonging to groups $\alpha$ and $\beta$ in layer $\ell$ exist with probability $q^{\ell}_{\alpha\beta}$. The observed adjacency matrix $A^{\O}$ is an aggregate that results from the combination of the links in each of the layers, but where all information of the layers has been lost (Fig.~\ref{f.schema}). We call this model the multilayer SBM.

Here we consider the simplest case of two layers, $L=2$. In such case, there are two combinations with a plausible physical interpretation: i) the {\it AND} combination of layers, in which $A^{\O}_{ij}=1$ if, and only if, $(i,j)$ are connected in both layers (Fig.~\ref{f.schema}(a)); ii) the {\it OR} combination of layers, in which $A^{\O}_{ij}=1$ if $(i,j)$ are connected in at least one layer (Fig.~\ref{f.schema}(b)). Indeed, each of these two mechanisms is plausible for specific scenarios. For example, the AND model is a plausible model for {\it in vivo} protein interactions, because in order for proteins to interact in the cell it is necessary for them to be capable of physically interacting (that is, to be linked in the layer of {\it in vitro} physical interactions) and to be expressed simultaneously in the same cellular compartment (that is, to be linked in the co-expression layer). The OR model is a plausible model for the effective on-line social network through which {\it memes} spread \cite{weng13}, 
because some people use Facebook to share memes, others use Twttier, and others use both.

In principle, we would like to identify which is the pair of partitions $(\P_1,\P_2)$ (in layers 1 and 2, respectively) that best describe the observed aggregate topology. The probabilistically complete way to solve this problem is to obtain the joint probability $P(\P_1, \P_2 |A^{\O})$ that $\P_1$ and $\P_2$ are the true partitions of the nodes given the aggregate observed network. This distribution is given by
\begin{eqnarray}
&& P(\P_1, \P_2 |A^{\O}) \propto  \label{eq.maxL} \\
&& \int DQ_{1}\int DQ_{2} \, P(A^{\O}| Q_1, Q_2, \P_1, \P_2) P(Q_1, Q_2, \P_1,\P_2) \nonumber
\end{eqnarray}
where $Q_\ell$ is a matrix whose elements $q^\ell_{\alpha\beta}$ represent the probability that a link exists between a pair of nodes belonging to groups $\alpha$ and $\beta$ in layer $\ell$, and $\int DQ_\ell \equiv\prod_{\alpha\leq \beta}\int_0^1dq^\ell_{\alpha\beta}$ is the integral over all possible values of these probabilities.

This integral can be computed both for AND combinations and for OR combinations of the two layers; for simplicity, here we focus on the AND model and discuss the OR model in the Appendices. Because in a SBM each links is independent of each other and in the AND model a link has to be present in both layers to appear in the observed aggregate network $A^{\O}$, the AND likelihood is
\begin{eqnarray}
&& P_{\rm AND}(A^{\O}| Q_1, Q_2, \P_1, \P_2) = \nonumber \\
%
%
&& = \prod_{ \alpha \leq \beta \brack \gamma \leq \delta }\left(q ^{1}_{\alpha\beta}q ^{2}_{\gamma\delta}\right)^{n^1_{\alpha\beta\gamma\delta}}  \left( 1- q ^{1}_{\alpha\beta}  
q ^{2}_{\gamma\delta}\right)^{n^0_{\alpha\beta\gamma\delta}},
\label{eq.and}
\end{eqnarray}
where
$n^1_{\alpha\beta\gamma\delta}$ is the number of links between pairs of nodes that are in groups $\alpha$ and $\beta$ respectively in layer 1, and in groups $\gamma$ and $\delta$ respectively in layer 2 ($n^1_{\alpha\beta\gamma\delta} = \sum_{i<j}A^{\O}_{ij}\delta_{\sigma^1_i\alpha}\delta_{\sigma^1_j\beta}\delta_{\sigma^2_i\gamma}\delta_{\sigma^2_j\delta}$); and $n^0_{\alpha\beta\gamma\delta}$ is the number of non-links between such pairs of nodes ($n^0_{\alpha\beta\gamma\delta} = \sum_{i<j}(1-A^{\O}_{ij})\delta_{\sigma^1_i\alpha}\delta_{\sigma^1_j\beta}\delta_{\sigma^2_i\gamma}\delta_{\sigma^2_j\delta}$).

Assuming a uniform distribution for the prior $P(Q_1, Q_2, \P_1,\P_2)={\rm const}$ \cite{guimera09} \footnote{Other possibilities include choosing non-uniform priors for the connection probabilities \cite{peixoto13,peixoto14,schmidt13} or different priors for the partitions \cite{decelle11,schmidt13,peixoto13,peixoto14,peixoto14b,yan14}.}, we can plug Eq.~(\ref{eq.and}) into Eq.~(\ref{eq.maxL}) and integrate to find (Appendices)
\begin{eqnarray}
 && P_{\rm AND}(\P_1, \P_2 |A^{\O}) \propto \label{eq.loglike} \\
 && \sum_{\{m_{rs}\} \brack m_{rs}=0,\dots,n^0_{rs}} \prod_{r,s} \frac{(-)^{m_{rs}}}{(n^1_{r}+m_{r}+1)(n^1_{s}+m_{s}+1)} {n^0_{rs} \choose m_{rs}} \label{eq.reall} \nonumber
 \end{eqnarray}
where, for clarity, we have used the shorthand $r\equiv \alpha\beta$ and $s\equiv \gamma\delta$, $m_r\equiv\sum_s m_{rs}$ and $m_s\equiv\sum_r m_{rs}$.

Given Eq.~(\ref{eq.loglike}), which is the complete probabilistic description of the multilayer SBM, one could in principle find the partitions $\P_1$ and $\P_2$ that maximize $P_{\rm AND}(\P_1, \P_2 |A^{\O})$. If this were possible, one would be able to perfectly disentangle the two SBMs responsible for the observed links, even though the observation did not have explicit information about the layers. It would also be possible to compare regular SBMs to multilayer SBMs to determine if a multilayer model is more or less appropriate to describe a given network. Unfortunately, the expression above becomes numerically intractable even for a small number of groups and therefore one needs to make approximations that simplify the problem.

\section{Link reliability with approximate multilayer stochastic block models}

\begin{figure}
  \includegraphics*[width=\columnwidth]{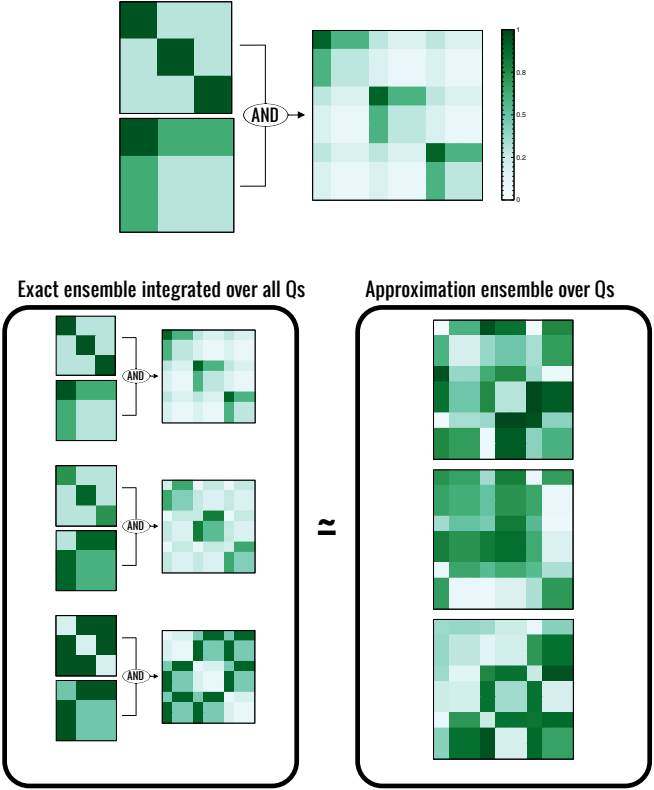}
  \caption{
  Exact and approximate multilayer SBM ensembles.
  {\bf (a)} Two independent single-layer SBMs aggregated using the AND mechanism. Each single-layer SBM is represented by its node-to-node connection probability matrix (represented by the shades of green; note that node ordering is different in each SMB). The aggregation of the two layers can also be represented as a single-layer SBM, in which each group comprises the nodes that belong to the same pair of groups $\alpha$ in layer 1 and $\beta$ in layer 2; this is the {\em intersection partition} $\P_I$. Moreover, if group $r$ in $\P_I$ corresponds to groups $\alpha$ in $\P_1$ and $\beta$ in $\P_2$, and group $s$ in $\P_I$ corresponds to groups $\gamma$ in $\P_1$ and $\delta$ in $\P_2$, then the probability of connection in the single-layer SBM is $q_{rs}^{\rm AND}=q_{\alpha\gamma}^1 q_{\beta\delta}^2$.
  {\bf (b)} For a fixed pair of partitions $\P_1$ and $\P_2$, we integrate over the ensemble of all possible probability matrices $Q_1$ and $Q_2$ (Eq.~(\ref{eq.loglike})). For each pair $(Q_1,Q_2)$, the resulting $q_{rs}^{\rm AND}$ are highly correlated. In our approximation, we assume that the elements $q_{rs}^{AND}$ are randomly drawn and independent of each other.
  \label{f.approx}
  }
\end{figure}

\begin{figure*}
  \includegraphics*[width=\textwidth]{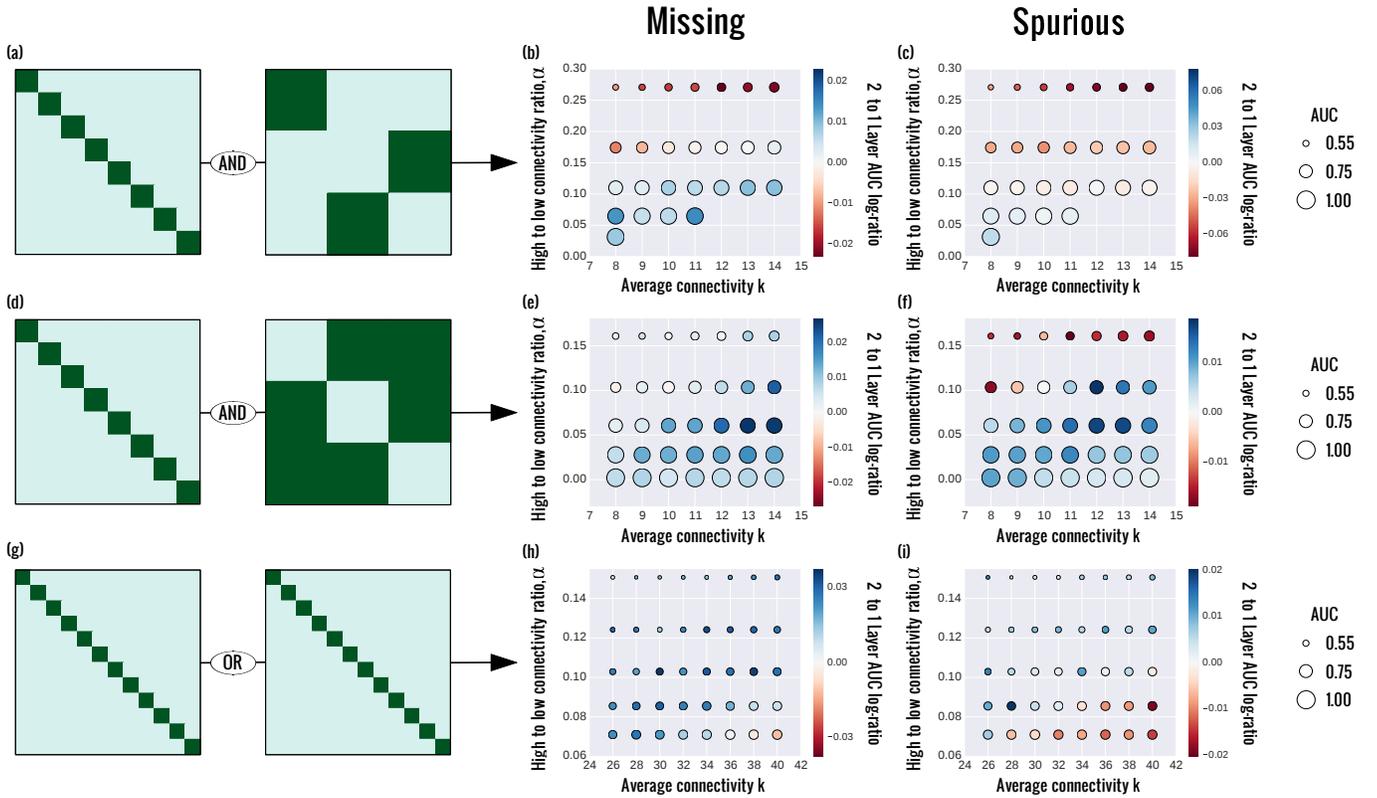}
  \caption{Performance of missing and spurious link identification on synthetic aggregated 2-layer networks.
  Each row corresponds to a different collection of 2-layer SBMs, which are illustrated in {\bf (a, d, g)}. Dark green corresponds to high connection probability $hp$ and light green to low connection probability $lp$, and we generate synthetic networks varying two parameters: the high-to-low connectivity ratio $\alpha=lp/hp<1$, and the average connectivity $k$ (Appendices). We compare the performance (AUC) at detecting missing links {\bf (b, e, h)} and spurious links {\bf (c, f, i)} of the approximate multilayer SBM approach, $AUC_{2L}$, against that of the the single-layer SBM approach, $AUC_{1L}$. The size of the circles represents the $AUC_{2L}$ of the multilayer approach. The color of the circles represents the logarithm of the ratio $\frac{AUC_{2L}}{AUC_{1L}}$, so that blue circles correspond to instances where the multilayer approach outperforms the single-layer approach, and conversely for red circles. 
\label{f.synthetic}}
\end{figure*}

We propose an approximation that makes it possible to work with multilayer SBMs. We start by noting that any multilayer SBM can be represented as a single-layer SBM (Fig.~\ref{f.approx}(a)) \footnote{The reverse is also true, so the possible network models one can generate with single-layer SBMs and multilayer SBMSs are, in fact, identical. However, it is important to note that each of them gives different weights to different models, so that a model that is relatively probable in the multilayer SBM family might be relatively rare in the single-layer SBM family, and vice versa.}. In the single-layer SBM, each group comprises the nodes that belong to the same pair of groups $\alpha$ in $\P_1$ and $\beta$ in $\P_2$ in the multilayer SBM (and only those); we call the single-layer partition the {\em intersection partition}. Moreover, if group $r$ in the intersection partition corresponds to groups $\alpha$ in $\P_1$ and $\beta$ in $\P_2$, and group $s$ in the intersection partition corresponds to groups $\gamma$ 
in $\P_1$ and $\delta$ in $\P_2$, then the probability of connection in the single-layer SBM is $q_{rs}^{\rm AND}=q_{\alpha\gamma}^1 q_{\beta\delta}^2$ (for simplicity, we again focus on the AND model and leave the OR model for the Appendices). This fully determines the single-layer SBM.

Here, we make the following approximation: we keep the information of the partitions $\P_1$ and $\P_2$ in the intersection partition, but consider that the matrix elements $q_{rs}^{\rm AND}$, while each being the result of the product of two factors,
are all independent of each other (see Fig.~\ref{f.approx}(b)). Since this approximation is equivalent to integrating separately every term with a different $(\alpha,\beta,\gamma,\delta)$ combination in Eq.~(\ref{eq.and}), it follows that the integrated likelihood depends exclusively on the intersection partition. In other words, within this approximation all pairs of partitions $(\P_1,\P_2)$ with the same intersection partition $\P_I$ are equally likely, and it is not possible anymore to uniquely determine the multilayer SBM that best describes the observed topology.

Despite this limitation, our approximation still enables us to address the fundamental question of whether real-world networks are better described by single-layer or multilayer models. Specifically, in what follows we compare the predictive power of single-layer and multilayer SBMs in the problem of detecting missing and spurious links in noisy networks \cite{guimera09}; we argue that, if (approximate) multilayer SBMs yield better predictions on real networks, then there is evidence to suggest that these networks are likely the outcome of multilayer processes (despite being observed as single-layer aggregates).

In the problem of assessing link reliability \cite{clauset08,guimera09}, the goal is to compute the probability $P(A_{ij}=1|A^{\O})$ that a link between nodes $i$ and $j$ truly exists ($A_{ij}=1$) given a noisy network observation $A^{\O}$, which contains false positives (spurious interactions that are reported but do not truly exist) and false negatives (missing interactions that truly exist but are not reported). We call the probability $R_{ij}=P(A_{ij}=1|A^{\O})$ the {\em reliability} of the link. In general, for any set $\M$ of models (single-layer SBMs, AND-multilayer SBMs or OR-multilayer SBMs), the reliability is \cite{guimera09}
\begin{eqnarray}
R^{\M}_{ij}=
 \frac{\int_{\cal M}dM P(A_{ij}=1| M) P(A^{\O}| M) P(M)}{Z},
\label{eq.inference}
\end{eqnarray}
where $Z$ is a normalization constant.
%

In the case of multilayer SBMs, the integral over the ensemble of models $\M$ requires: i) the integration over the connection probabilities $Q_1$ and $Q_2$ (akin to what we did to obtain Eq.~(\ref{eq.maxL})); ii) the sum over all pairs of partitions $\P_1$ and $\P_2$.
Within our approximation, the first step can be carried out analytically but the second cannot (Appendices). However, always within our approximation, one can exploit the fact that the integral in Eq.~(\ref{eq.inference}) depends exclusively on the intersection partition $\P_I$ and map the sum over pairs of partitions onto a sum over a single partition.
By doing so we obtain the following expression for the link reliability (see Appendices for the analogous expression for the OR model)
\begin{eqnarray}
&& R^{\rm AND}_{ij}= \label{eq.inf1}\\
&& \frac{1}{Z} \sum_{{\P}_I} \Bigg( \frac{n^1_{\sigma_i\sigma_j}+1}{n_{\sigma_i\sigma_j}+2} \cdot 
\frac{\sum_{k=n^1_{\sigma_i\sigma_j}+2}^{n_{\sigma_i\sigma_j}+2} \frac{1}{k} }{ \sum_{k=n^1_{\sigma_i\sigma_j}+1}^{n_{\sigma_i\sigma_j}+1} \frac{1}{k} } \cdot D({\P}_I)\cdot
e^{-\mathcal{H}({\P}_I)} \Bigg) \nonumber
\end{eqnarray}
where the sum is over all possible intersection partitions (that is, all single-level partitions), $n^1_{\alpha\beta}$ is the number of links between groups $\alpha$ and $\beta$ in the intersection partition, $n_{\alpha\beta}=n^0_{\alpha\beta}+n^1_{\alpha\beta}$ is the number of pairs of nodes in groups $\alpha$ and $\beta$, and $D({\P}_I)$ is the number of pairs $(\P_1,\P_2)$ that have the same intersection partition $\P_I$ (see Appendices). The energy function $\mathcal{H}$ is
\begin{eqnarray}
&& \mathcal{H}({\P}_I) = \label{eq.energy} \\
&& \sum_{\alpha \leq \beta} 
\Bigg(\ln (n_{\alpha\beta}+1) +\ln{n_{\alpha\beta} \choose n^1_{\alpha\beta}} - \ln \bigg( \sum_{k=n^1_{\alpha\beta}+1}^{n_{\alpha\beta}+1} \frac{1}{k} \bigg) \Bigg) \nonumber
\end{eqnarray}
where the sum is over all distinct pairs of groups in $\P_I$.

As in \cite{guimera09}, the expression for the link reliability (Eq.~\ref{eq.inf1}) is analogous to an ensemble average of an observable in statistical mechanics, giving $\mathcal{H}({\cal P}_I)$ the meaning of an energy associated to a specific intersection partition. We can use a Markov Chain Monte Carlo algorithm to compute numerically $R_{ij}$ (see Supplementary Material for details). As it turns out, $\mathcal{H}({\cal P}_I)$  is equal to the energy obtained assuming a single SBM (Eq. S2, \cite{guimera09}) plus a term that arises because of the fact that the probability matrix elements associated 
to the intersection SBM are the result of a product of two probabilities. In a Bayesian context, we can interpret this term and the degeneration $D(\P_i)$ as non-uniform priors for the intersection partitions.

\section{Validation of link reliability estimation in model networks}

Now that we are able to estimate link reliabilities using single-layer SBMs \cite{guimera09} and our approximation to two-layer (AND and OR) SBMs (Eq.~(\ref{eq.inf1})), we compare the performance of these approaches at detecting missing and spurious interactions. Our expectation is that if real-world networks are truly the result of the aggregation of multiple layers, assuming a two layer structure should result in higher accuracy. 

To identify the limits of detectability of the 2-layer SBM model, we first construct a set of multilayer test networks that have a clearly differentiated block structure in each of two layers, and that are aggregated using the AND and OR models (see Methods and Fig.~\ref{f.synthetic}).
We consider the predictive power of each of the approaches at detecting \cite{clauset08,guimera09}: i) missing links (we remove a fraction $f$ of the links and compute the fraction of times that a removed link has a higher reliability than a link not present in the original network, that is the AUC statistic); ii) spurious links (we add a fraction $f$ of links and compute the fraction of times that an added link has a lower reliability than a link present in the original network, that is the AUC statistic).

For AND networks (Fig.~\ref{f.synthetic}(a-f)) we find that, for the detection of both missing and spurious links, the 2-layer approach outperforms the single-layer approach, especially: (i) when the number of distinct node groups in the intersection partition and the connectivity grow; (ii) for small or moderate noise levels (fraction of removed/added links. Only when the structure of the blocks becomes very blurry do we observe that the single-layer approach works better (but in this region all approaches do in fact work poorly).

For OR networks (Fig.~\ref{f.synthetic}(g-i)), the 2-layer approach again outperforms its single-layer counterpart in most situations. In this case, however, the largest improvements in performance happen for the hard cases with lower connectivity. This can be explained by noting that 
%
the OR model tends to generate very dense networks, whereas aggregate AND networks are sparser than the networks in each of the layers.
%
%
Therefore, in general we expect the AND model to produce better results in real-world networks.

\section{Multilayer stochastic block models are more predictive for real networks}

\begin{figure}
  \includegraphics*[width=\columnwidth]{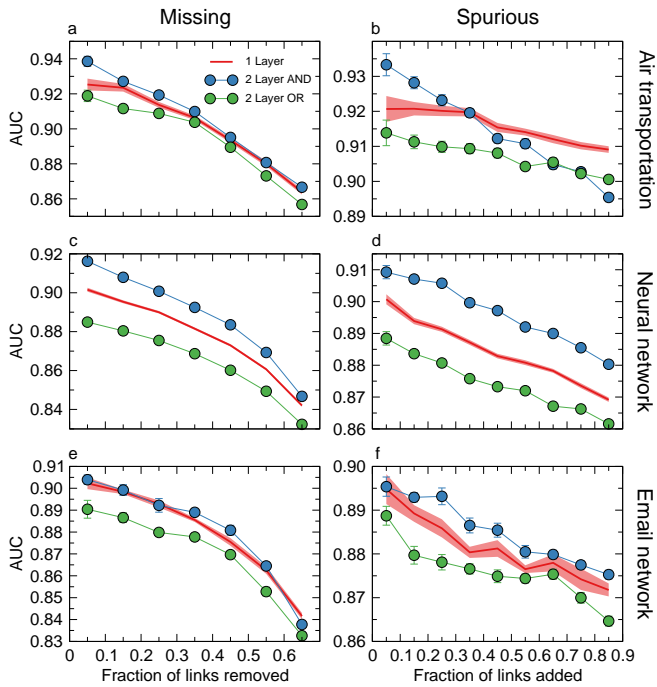}
  \caption{Performance of missing and spurious link identification on real networks.
  To compare the performance of the different approaches at detecting missing links {\bf (a, c, e)}, we randomly remove a fraction of the links (false negatives) from the real network and calculate the reliability of each unobserved link. Then we rank the links by decreasing score and calculate how often a removed link (false negative) has a higher reliability that a link that is truly non-existent in the real network (true negative). Analogously, to detect spurious links {\bf (b, d, f)} we randomly add a fraction of links (false positives), calculate the reliability of the observed links, and calculate how often an added link (false positive) has a lower reliability that a link that is truly existent in the real network (true positive).
  We tested the analysis on three real networks: {\bf (a, b)} the air transportation network in Eastern Europe; {\bf (c, d)} the neural network of {\textit (C. elegans)}; {\bf (e, f)} the email network within an organization.
  \label{f.real}}
\end{figure}

Finally, we compare the performance of the single-layer and multilayer approaches on three real-world networks: (i) the air transportation network in Eastern Europe \cite{guimera05b}; (ii) the neural network of {\it C. elegans} \cite{white86}; and (iii) the email network within a university \cite{guimera03}. Our results show that the two-layer AND model provides a better description of these real-world networks since both missing and spurious interactions are more accurately detected by the multilayer SBM approach consistently (the improvement is slight but, in most cases, significant), especially for low observational noise.

\section{Discussion}

We have introduced the family of multilayer SBMs, which generalizes single-layer SBMs to situations where links arise in different layers and are aggregated through different mechanisms. We have also given the probabilistically complete solution to the problem of inferring the optimal multilayer SBM for a given aggregate network, and proposed a tractable approximation which enables us to objectively address the question of whether an observed network is best described as the projection of multiple layers or as a single layer. Our results suggest that many real-world networks are indeed projections.

Although, as mentioned above, there have been proposals to extend the concept of modularity to multilayer networks~\cite{mucha10}, ours represents a pioneering attempt to extend stochastic block models to multilayer systems. In this regard, it is important to stress that in this work we are concerned with the learning of multilayer models from aggregate networks where all information about the layers has been lost; in this sense, our work is different from previous attempts to do inference of stochastic block models on multigraphs where the layers themselves are observed \cite{guimera12}.

Our work is also different from works on link prediction using latent feature models \cite{miller09,palla12,kim13}. An important difference between latent feature approaches and ours is that the latent feature model considers that the probability of existence of a link is a function of the weighted sum of the interactions at the different layers; therefore, the latent feature model does not allow a physical interpretation of what each layer is and of how layers are combined. All in all, latent feature models are very well suited for the inference of unobserved links, but due to the intricacies of the model and the difficulty to interpret its ``parameters,'' it is not clear whether they are appropriate to address the question of whether a real network is really the outcome of a multilayer process or not (and may it may also be prone to overfitting when observational data is noisy).

Our multilayer SBM is the simplest group-based multilayer model one can propose. We believe that its detailed analysis will open the door to better understand the structure of real complex networks.

\begin{acknowledgments}
We thank the following people for helpful comments and discussions: A. Aguilar-Mogas, A. Arenas, M. De Domenico, A. Godoy-Lorite, T.P. Peixoto, O. Senan-Campos, M. Tarr\'es-Deulofeu. This work was supported by a James S. McDonnell Foundation Research Award, Spanish Ministerio de Economía y Comptetitividad (MINECO) Grant FIS2010-18639, European Union Grant PIRG-GA-2010-277166 (to RG), European Union Grant PIRG-GA-2010-268342 (to MSP), and European Union FET Grant 317532 (MULTIPLEX).
\end{acknowledgments}

\bibliography{../References/ref-database,ref-marta}

\end{document}